# Deep Learning-based Protoacoustic Signal Denoising for Proton Range Verification

Jing Wang[1], James J. Sohn[2], Yang Lei[1], Wei Nie[3], Jun Zhou[1], Stephen Avery[4], Tian Liu[5*] and Xiaofeng Yang[1*]

[1]Department of Radiation Oncology and Winship Cancer Institute, Emory University, Atlanta, GA
[2]Department of Radiation Oncology, Northwestern University, Chicago IL
[3]Department of Radiation Oncology, University of Nebraska, Omaha, NE
[4]Department of Radiation Oncology, University of Pennsylvania, Philadelphia, PA
[5]Department of Radiation Oncology, Mount Sinai Medical Center, New York, NY 10029
*Email: xiaofeng.yang@emory.edu and tian.liu@mountsinai.org

**Running title:** Protoacoustic Signal Denoising

**Manuscript Type:** Original Research

## Abstract

**Objective**: Proton therapy offers an advantageous dose distribution compared to the photon therapy, since it deposits most of the energy at the end of range, namely the Bragg peak (BP). Protoacoustic technique was developed to *in vivo* determine the BP locations. However, it requires large dose delivery to the tissue to obtain an averaged acoustic signal with a sufficient signal to noise ratio (SNR), which is not suitable in clinics. We propose a deep learning-based technique to acquire denoised acoustic signals and reduce BP range uncertainty with much lower doses.

**Approach:** Three accelerometers were placed on the distal surface of a cylindrical polyethylene (PE) phantom to collect protoacoustic signals. In total 512 raw signals were collected at each device. Device-specific stack autoencoder (SAE) denoising models were trained to denoise the input signals, which were generated by averaging 1, 2, 4, 8, 16, or 32 raw signals. Both supervised and unsupervised learning training strategies were tested for comparison. Mean squared error (MSE), signal-to-noise ratio (SNR) and the Bragg peak (BP) range uncertainty were used for model evaluation.

**Main results:** After SAE denoising, the MSE was substantially reduced, and the SNR was enhanced. Overall, the supervised SAEs outperformed the unsupervised SAEs in BP range verification. For the high accuracy detector, it achieved a BP range uncertainty of $0.20 \pm 3.44$ mm by averaging over 8 raw signals, while for the other two low accuracy detectors, they achieved the BP uncertainty of $1.44 \pm 6.45$ mm and $-0.23 \pm 4.88$ mm by averaging 16 raw signals, respectively.

**Significance:** We have proposed a deep learning based denoising method to enhance the SNR of protoacoustic measurements and improve the accuracy in BP range verification, which greatly reduces the dose and time for potential clinical applications.

## 1. Introduction

Proton therapy is an ion therapy that receives increasing interest in both research studies and clinical applications in radiation oncology. The linear energy transfer (LET) of proton dramatically increases for small velocities and thus most of the particle energy is deposited at the end of its trajectory before completely stopped in materials, resulting a so-called Bragg peak (BP) (Newhauser and Zhang, 2015; Mohan and Grosshans, 2017). Due to the large LET at the end of range, none or minimal doses are deposited beyond the range (Wilson, 1946). Compared with the photon therapy, whose dose distribution decreases exponentially in deep tissues, the proton therapy provides a well-characterized dose depth and a more conformal dose. Such good conformity can benefit to accurately deliver a high dose to tumors, while largely sparing the adjacent organs at risk (OARs). Therefore, the range verification of BP within tissues is critical in treatment. Several noninvasive techniques including the positron emission tomography (PET) (Parodi *et al.*, 2007; Parodi, 2015) and prompt gamma imaging (Min *et al.*, 2006) have been proposed to localize the BP location *in vivo* during real-time radiation therapy (Knopf and Lomax, 2013). However, both methods depend on bulky and complex instrumentation and only reveal indirect information of the BP position, and hard to achieve accuracy in a few millimeters for clinical application.

Protoacoustic determination of BP range has been actively studied for the localization of BP since a direct correlation between the acoustic signals and BP locations can be utilized (Sulak *et al.*, 1979; Hayakawa *et al.*, 1995; Albul *et al.*, 2001; Bychkov *et al.*, 2008; Jones *et al.*, 2016). When pulsed proton beam deposits energy in medium, the energy would dissipate in fast heat expansion and emit acoustic waves due to thermoacoustic conversion of heat to pressure. It is proposed that the proton range verification can be directly measured based on the time-of-flight (TOF) of the generated acoustic waves. Another advantage of the protoacoustic technique is that it may be used to monitor proton dose distribution in a patient in real-time owing to its relatively simple device-setup compared to PET or prompt gamma imaging. However, it remains a challenging task since proton acoustic signal is very weak and noisy, and it typically requires delivering a large number of pulses to a single spot in the medium to obtain high-quality proton acoustic signals with a high signal-to-noise ratio (SNR). Jones *et al.* (Jones *et al.*, 2015) averaged as high as 2048 single pulses to obtain a stable and accurate measurements, Nie *et al.* (Nie *et al.*, 2018) averaged over 1024 signals, and Kellnberger *et al.* (Kellnberger *et al.*, 2016) used a lower number of 512 averages for their 3D ionoacoustic scans. The large number of signals required to average for the final measurement will result in high doses (*e.g.*, averaging over 2048 proton pulses (Jones *et al.*, 2015) correspond to 38.9 Gy ) to the medium as well as a long beam delivery time, hindering the clinical application of protoacoustic range verification. If BP can be identified with as few measurements as possible, both the delivery dose and time can be reduced, and that is the way forward to the clinic. Therefore, an appropriate denoising method to improve the acoustic signal acquisition is essential to apply protoacoustic technique in clinics.

In previous literatures, denoising techniques such as low pass filter (Freijo *et al.*, 2021) and wavelet-based transformation (Sohn *et al.*, 2020) were employed to eliminate the high-frequency noise. However, such methods rely on the choice of various thresholding methods such as hard and soft thresholding. Moreover, though a large portion of the noise is removed in the reconstructed signal, the residual noise, which is often complex with an unknown distribution in the frequency domain, needs to be treated by advanced approach. Recently, data-driven approaches with deep learning-based methods have demonstrated success in denoising applications, specifically, the stacked autoencoder (SAE) paradigm for local denoising and feature learning (Vincent *et al.*, 2010; Vincent *et al.*, 2008). SAE composes of an encoder to learn the useful higher-level representations and a decoder to reconstruct the denoised signal, thus removing the noisy components in the original corrupted input. In medicine field, SAEs have been widely used in the electrocardiogram (ECG) denoising and significantly enhanced the signal to noise ratios (Nurmaini *et al.*, 2020; Xiong *et al.*, 2016; Xiong *et al.*, 2015; Liu *et al.*, 2021). Our hypothesis is that the proton acoustic signals, which are corrupted with noises in the similar pattern as the ECG signals, could also be denoised with the SAE network while keeping as few measurements as possible. In addition, a patch-based method was utilized for data augmentation to address overfitting. The long signals were first cut into smaller sections with moving origins and large overlaps, and then the small sections were fed into the SAEs for denoised output. Finally, the denoised small patches were put back to their corresponding origins and a long-merged signal was obtained by averaging these small overlapping sections. In this work, we used three detectors to collect numerous protoacoustic signals generated by proton pulses in a plastic phantom, and then utilized SAEs to denoise the proton acoustic signals while preserving the BP signal with minimized signal acquisitions.

## 2. Method

### 2.1 Experiment setup and signal acquisition

To collect protoacoustic signals, three detectors (accelerometers) were placed on the distal surface of a cylindrical polyethylene (PE) phantom (diameter=20.88 cm, length=33.58 cm) as illustrated in Figure 1 (a). A 226 MeV proton beam was first attenuated by a 2 cm solid water (Gammex 457-CTG, Middleton, WI, USA), and then the proton pulse was incident onto the PE phantom from one end and generated a BP inside the phantom, emitting protoacoustic signals. The acoustic signals were measured by the three detectors placed on the other end. The three detectors are two accelerometers of relatively low accuracy (Type 4374, Denmark) and one accelerometer of high accuracy (Brüel&Kjær, Type-4017-C), namely L1, L2, and H. The electric charge signal was amplified (65 dB, ×1780 gain) and filtered (10 Hz high pass and 100 kHz low pass filter) by a Nexus charge amplifier 2692 before output through a 4-channel digital oscilloscope (Picoscope 5444B, PicoTech, UK).

In total 512 proton pulses were incident onto the PE phantom, and each proton pulse would produce a BP inside the phantom. An incident proton pulse has an average current of 490 nA, consisting of $5.7 \times 10^7$ protons, which is equivalent to 2.36 cGy of dose delivery to the BP in phantom. For one BP, each of the three detectors would independently collect an acoustic signal emitted from the BP, which is to say, we collected altogether 3*512 raw acoustic signals with three detectors and each detector collected 512 raw data. By averaging the 512 single raw waves, we can obtain a clean and stable signal for each detector. The averaged signals of three detectors as well as the proton beam are shown in Figure 1(b). For easy comparison, all signals have been normalized to [0,1].

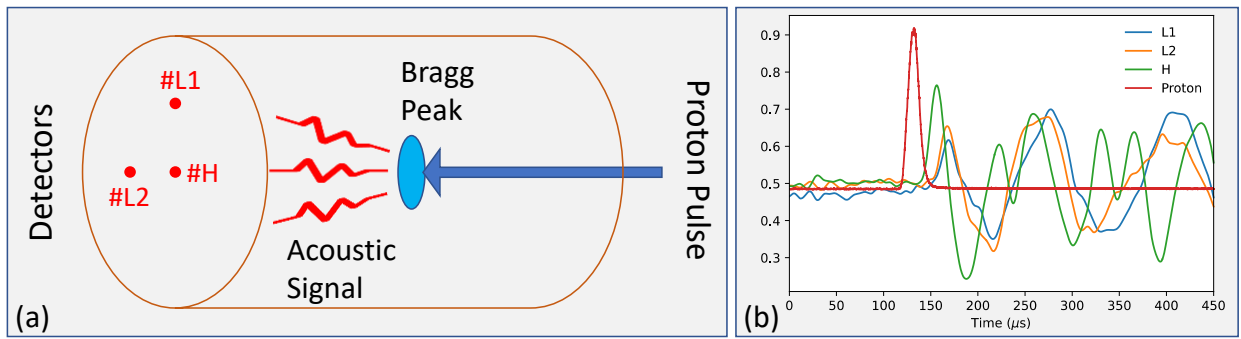

**Figure 1**. (a) Experimental setup of measuring protoacoustic signals. A 226 MeV proton beam entered the PE phantom and produced a Bragg peak (BP) inside the medium, emitting pressure waves due to thermal expansion. The acoustic

signals were collected by three detectors at the other end of PE phantom. (b) The clean signals of three detectors (L1, L2, and H) and proton pulses averaged over 512 measurements.

## 2.2 BP range verification

Protoacoustic is a straightforward technique to measure the location of BP directly from the time-of-fly (TOF) between proton pulses and arrival of acoustic waves. As shown in Figure 2, the TOF is characterized by the time elapse between the minimum of proton pulse and the first maximum of acoustic wave. The BP position can be calculated through the equation below:

$$\tau_i = \frac{d_i}{c} + \delta_i,$$

where $\tau_i$ is the acoustic TOF for the $i^{th}$ detector, $c$ is the speed of sound in PE phantom (2.07 mm/μs), $d_i$ is the distance between BP location and the $i^{th}$ detector, and $\delta_i$ is the unique responsive time of each detector. $\delta_i$ can be calibrated through systematic phantom experiments as described in previous studies (Nie *et al.*, 2018), but the calibration of detector responsive time is not of interest in this work. Our goal is to locate the first maximum of acoustic wave that characterizes the arrival of acoustic signal, since the identification of the acoustic arrival from a noisy signal is one of the most difficult tasks in reducing the number of pulses needed for protoacoustic measurements, and yet potentially feasible with SAE reconstruction method, as we will show in the following sections.

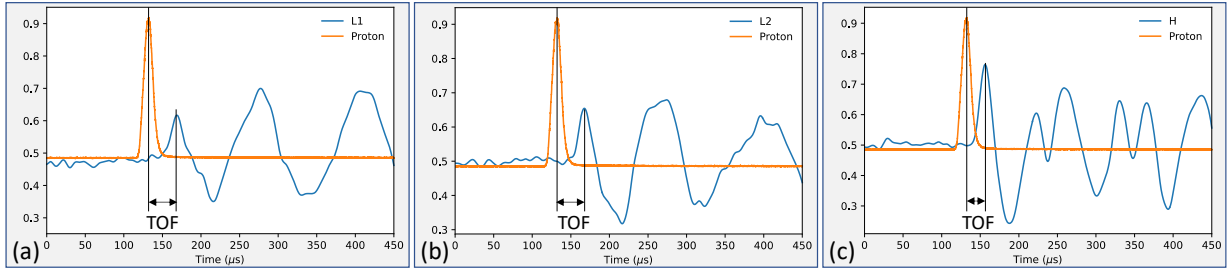

**Figure 2**. (a) The L1 and proton signals averaged over 512 raw waves. Time-of-flight (TOF) can be measured from the time elapse between proton pulse signal peak and the arrival of acoustic signal (the first peak of the pressure wave). (b) and (c) show the TOFs of L2 and H detectors, respectively.

## 2.3 Workflow and training strategy

After we have obtained the 512 raw signals for each detector (L1, L2, and H), SAE networks were trained to reconstruct the denoised signals from fewer number of raw signals. Since each detector is a unique device, e.g., has specific response time, we trained device-specific models for each detector. For each detector, we randomly chose 256 raw signals for model training and the remaining 256 raw signal for external testing.

For the training dataset, the grand ground truth (GT) is obtained by averaging over the 256 raw signals. We then generated the input noisy signals by averaging a small number (e.g., 1, 2, 4, 8, 16, and 24) of raw waves, and generated the corresponding clean signals by averaging 192 out of the 256 raw waves, including the raw waves used in the noisy signals. We employed two training strategies to learn denoising models using SAE. The first strategy is a supervised learning that trains the input noisy signals with the clean signals as learning target, and the other strategy is a self-supervised learning that minimizes the reconstruction error between the input noisy signals and output denoised signals, without any reference to the generated clean signals (unsupervised learning) (Vincent *et al.*, 2010). These two training strategies are named as SAE_clean and SAE_self, respectively. During the training phase, for SAE_clean, the input are noisy signals and ground truth are clean signals for the model, while for SAE_self, the input are noisy signals but the target ground truth is also the input noisy signals. The differences between the two training strategies would be reflected in the loss function, as discussed in the following section. For the other 256 raw signals in testing dataset, similar treatment was applied to obtain the grand GT signal, the noisy test signals, and clean targe signals. After training the SAEs, a new noisy signal from the test dataset was fed into the trained SAE model to obtain the denoised signal, and subsequently compared to the grand GT for model performance evaluation.

Due to limited resources, we only performed the protoacoustic measurements at one energy level of 226 MeV, so the patterns of acoustic waves are highly limited, basically containing only three intrinsically independent patterns since we had three detectors. If we only consider the whole long signals (more than 450 µs) as training objectives, it is prone to overfitting. To address the overfitting issue, our SAEs were trained in a patch-based manner. By cutting the long signals into much smaller sections with shifting origins, we were able to greatly augment the size and variation of training and testing dataset. The reconstructed long signals were obtained by merging the SAE output patches. The workflow of the denoising the acoustic signals is sketched in Figure 3.

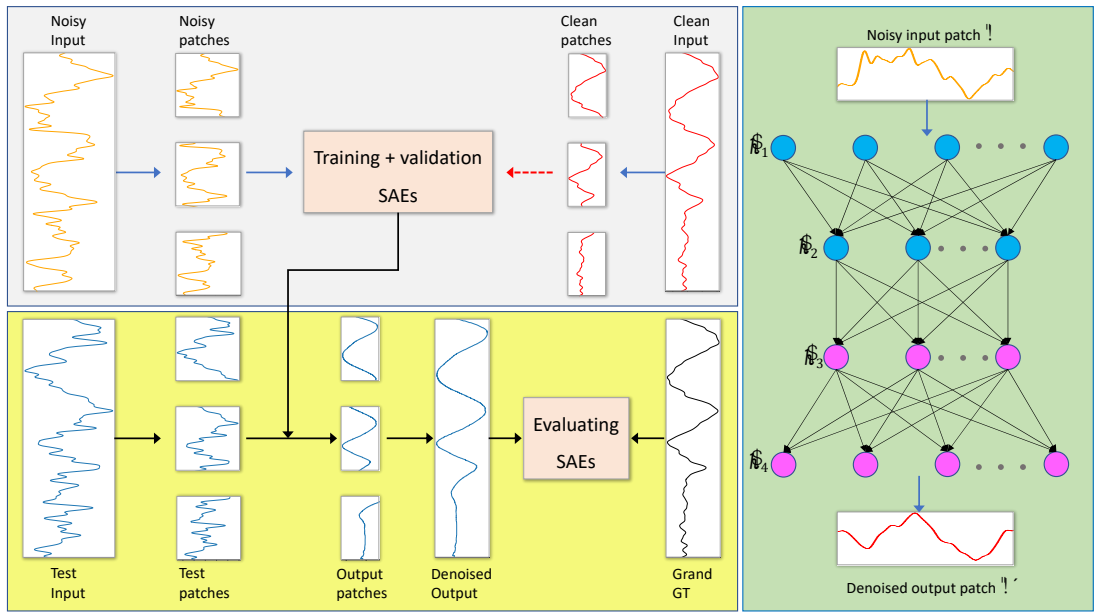

**Figure 3**. Workflow of denoising protoacoustic signals and the sketched diagram of SAEs.

## 2.4 Network architectures

The architectures of the proposed SAEs are shown in Fig. 3, composing of a compressing encoder path and an expanding decoding branch. All the one-dimensional input noisy signals are first normalized into [0,1] and cut into small patches as $\hat{x}$. Similarly, the patches of corresponding training clean signals $\hat{y}$ were also obtained. Through encoder layers, the input $\hat{x}$ is encoded to output $\hat{h}$ in the hidden layers. This step serves as extracting higher-level useful representations of the signals and eliminating the redundant noisy components of the input. Then, the useful latent features $\hat{h}$ expand to the output $\hat{x}'$ via decoder layers. After that, the aim of SAEs is to minimize the reconstruction error either between $\hat{x}'$ and $\hat{y}$ for SAE_clean, or between $\hat{x}'$ and $\hat{x}$ for SAE_self strategy.

Denoting $\theta_i = \{w_i, b_i\}$ are the weighting and bias parameters for the $i_{th}$ SAE layer and $s(x) = 1/(1 + exp(-x))$ is a sigmoid activation function to introduce nonlinearity between SAE layers, the hidden output can be represented as

$$\hat{h}_1 = s(w_1 \hat{x} + b_1),$$

$$\hat{h}_i = s(w_i' \hat{h}_{i-1} + b_i), \quad i = 2, 3, 4.$$

Denoting $\theta' = \{w_{out}, b_{out}\}$ are weighting and bias parameters for the last decoder layer, we define the decoder's output as

$$\hat{x}' = s(w_{out} \hat{h}_4 + b_{out}).$$

For each denoised signal $\hat{x}'$, it is reconstructed form the noisy signal $\hat{x}$, and is expected to resemble either the clean signals $\hat{y}$ (SAE_clean) or the input $\hat{x}$ (SAE_self). Thus, for SAE_clean the learnable parameters are optimized by minimizing the reconstruction error:

$$\theta_i, \theta' = arg \min_{\theta_i, \theta'} \|\hat{x}' - \hat{y}\|_2^2, or$$

for SAE_self the learnable parameters are optimized by minimizing the reconstruction error:

$$\theta_i, \theta' = arg \min_{\theta_i, \theta'} \|\hat{x}' - \hat{x}\|_2^2.$$

where the *L2* norm aims to minimize the difference between $\hat{x}'$ and the learning target.

### 2.5 Experiment design and training parameters

In order to arrange all of the experiments mentioned above, we had to perform multiple experiments for both SAE_clean and SAE_self strategies, the different numbers of averaging (n_avg) input raw signals, as well as the three detectors. Thus, we name the series of experiments as: clean/self- detector name-n_avg. For example, "clean-L1-2" means the experiment used pressure waves collected on detector L1 (one of the low accuracy detectors), averaged 2 raw signals to obtain the noisy input, and employed the SAE_clean strategy, while "self-H-16" means the experiment used pressure waves collected on detector H (the high accuracy detector), averaged 16 raw signals as the noisy input, and employed the SAE_self strategy to train the SAE model. During the training phase, 100 long noisy input signals were generated by randomly averaging 1, 2, 4, 8, 16, or 24 raw signals, and the patch size used for data augmentation was 66 µs with an origin shift of 3.2 µs. After data augmentation, we obtained 13400 signal patches for training. Among the 13400 patches, 2/3 of the data was used for training and 1/3 was used for validation of SAE models.

### 2.6 Model evaluation and BP range verification

Two metrics were used for performance measurement, mean squared error (MSE) and SNR:

$$MSE = \frac{1}{n} \sum_{i=1}^{n} (s_i - x_i)^2,$$

$$SNR = 10 \log_{10} \frac{\sum_{i=1}^{n} x_i^2}{\sum_{i=1}^{n}(s_i - x_i)^2}, \tag{5}$$

Where $s_i$ and $x_i$ are the long grand GT signal and long noisy/denoised signals from the test dataset (the remaining 256 raw signals outside the training dataset), respectively. Both MSE and SNR were computed for noisy and denoised signals, to evaluate the efficacy of SAEs.

Moreover, we evaluated the BP displacement between the pre-training and after-training signals by calculating the time differences based on the grand GT. As we have discussed in previous sections, in this work, we only focus on identifying the first maximum of pressure wave, which characterize the arrival of acoustic signal. We used the first peak of grand GT (averaging over all 256 raw signals) as the reference timepoint of acoustic arrival. For simplicity, we automatically located the maximum between 0 to 192 µs of a merged long acoustic signal to identify the arrival of first pressure peak. For both noisy and denoised signals, their first peaks can have shifts from the reference timepoint. The time shift of TOF multiplied by the sound speed (about 2.07 mm/µs in the experimented PE phantom) corresponds to the uncertainty of BP range verification. To estimate the BP range uncertainty before and after denoising, we used two metrics to calculate the shifts, mean error of the shifts ($ME^{BP}$) and mean absolute error of the shifts ($MAE^{BP}$). This step is helpful to intuitively examine the feasibility of applying the SAE denoising in clinics.

## 3. Results

### 3.1 Evaluation of SAEs on the test dataset

Table 1 shows the MSE and SNR values of L1 detector signals before and after denoising on the test dataset, using the SAE_clean training strategy. The metrics are calculated from the first 400 µs of merged long signals. Table 2 shows the MSE and SNR changes in percentage for all experiments on the test set. Figure 4 shows the example noisy input signals, reconstructed denoised signals, and the grand GT signal of H detector, for both SAE_clean and SAE_self using various n_avg. For both training strategies, the MSE values of after-denoising were dramatically reduced while the SNR were enhanced. However, at low n_avg (e.g., 1, 2, and 4), SAE_clean outperformed SAE_self in both metrics.

**Table 1**. MSE and SNR values of L1 detector signals before and after denoising on the test dataset (SAE_clean strategy).

| n_avg | MSE_before | MSE_after | MSE %change | SNR_before(dB) | SNR_after(dB) | SNR %change |
|---|---|---|---|---|---|---|
| 1 | 3.39E-02 | 5.29E-03 | -84.38% | 9.08 | 16.99 | 87.21% |
| 2 | 1.62E-02 | 3.01E-03 | -81.41% | 12.06 | 18.98 | 57.40% |
| 4 | 8.00E-03 | 1.92E-03 | -76.03% | 15.03 | 21.02 | 39.87% |
| 8 | 3.81E-03 | 9.62E-04 | -74.74% | 18.19 | 24.19 | 33.03% |
| 16 | 1.85E-03 | 4.93E-04 | -73.28% | 21.24 | 26.99 | 27.04% |
| 24 | 1.22E-03 | 3.88E-04 | -68.24% | 27.96 | 27.96 | 21.22% |

**Table 2**. MSE and SNR changes in percentage after denoising for all experiments on the test set.

| Experiment# | MSE | SNR | Experiment# | MSE | SNR |
|---|---|---|---|---|---|
| clean-L1-1 | -84.38% | 87.21% | self-L1-1 | -28.36% | 14.81% |
| clean-L1-2 | -81.41% | 57.40% | self-L1-2 | -36.09% | 13.46% |
| clean-L1-4 | -76.03% | 39.87% | self-L1-4 | -52.69% | 20.51% |
| clean-L1-8 | -74.74% | 33.03% | self-L1-8 | -52.57% | 17.75% |
| clean-L1-16 | -73.28% | 27.04% | self-L1-16 | -57.61% | 18.10% |
| clean-L1-24 | -68.24% | 21.22% | self-L1-24 | -68.03% | 21.80% |
| clean-L2-1 | -85.07% | 85.44% | self -L2-1 | -32.46% | 14.37% |
| clean-L2-2 | -78.49% | 53.24% | self -L2-2 | -38.92% | 16.20% |
| clean-L2-4 | -74.35% | 39.40% | self -L2-4 | -50.61% | 21.19% |
| clean-L2-8 | -44.37% | 16.33% | self -L2-8 | -63.73% | 25.10% |
| clean-L2-16 | -58.56% | 18.75% | self -L2-16 | -63.80% | 21.18% |
| clean-L2-24 | -50.58% | 13.09% | self -L2-24 | -57.27% | 16.47% |
| clean-H-1 | -79.46% | 59.03% | self -H-1 | -28.13% | 12.99% |
| clean-H-2 | -71.09% | 41.85% | self -H-2 | -33.34% | 15.04% |
| clean-H-4 | -64.60% | 26.40% | self -H-4 | -53.61% | 18.86% |
| clean-H-8 | -65.14% | 23.29% | self -H-8 | -60.09% | 20.67% |
| clean-H-16 | -67.16% | 21.82% | self -H-16 | -66.10% | 21.72% |
| clean-H-24 | -46.59% | 10.48% | self -H-24 | -60.15% | 16.43% |

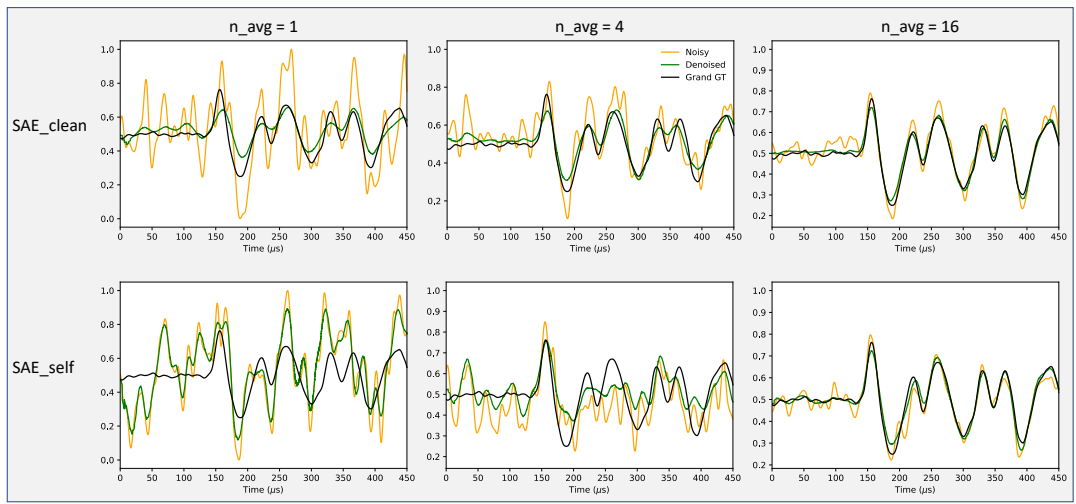

**Figure 4**. Examples of the noisy input signals (orange curves), reconstructed denoised signals (green curves), and grand GT signals (black curves) obtained on the high accuracy detector.

### 3.2 BP range uncertainty

From our experiments, the BP uncertainties began to show obvious improvement at n_avg greater than 2 for the detector H and at n_avg greater than 4 for detector L1 and L2, especially considering the largely decreased standard deviation of $ME^{BP}$ and $MAE^{BP}$. Table 3 shows the $ME^{BP}$ and $MAE^{BP}$ representing BP range verification uncertainties before and after denoising, with n_avg starting at 4. The values have been converted to mm by multiplying the TOF peak shifts with the sound speed in PE phantom (2.07 mm/μs). Figure 5 shows examples of enhanced BP locations after SAE_clean denoising with n_avg of 4, 8 and 16, on all the three detectors. For better observe the correction of peak localization, we zoom in the range between 100 and 192 μs in Figure 5. Among the three detectors, the high accuracy detector H is most accurate in identifying BP positions, with the least range uncertainties. Besides, with the increasing n_avg of raw signals in the input, the BP range verification got improved both before and after denoising.

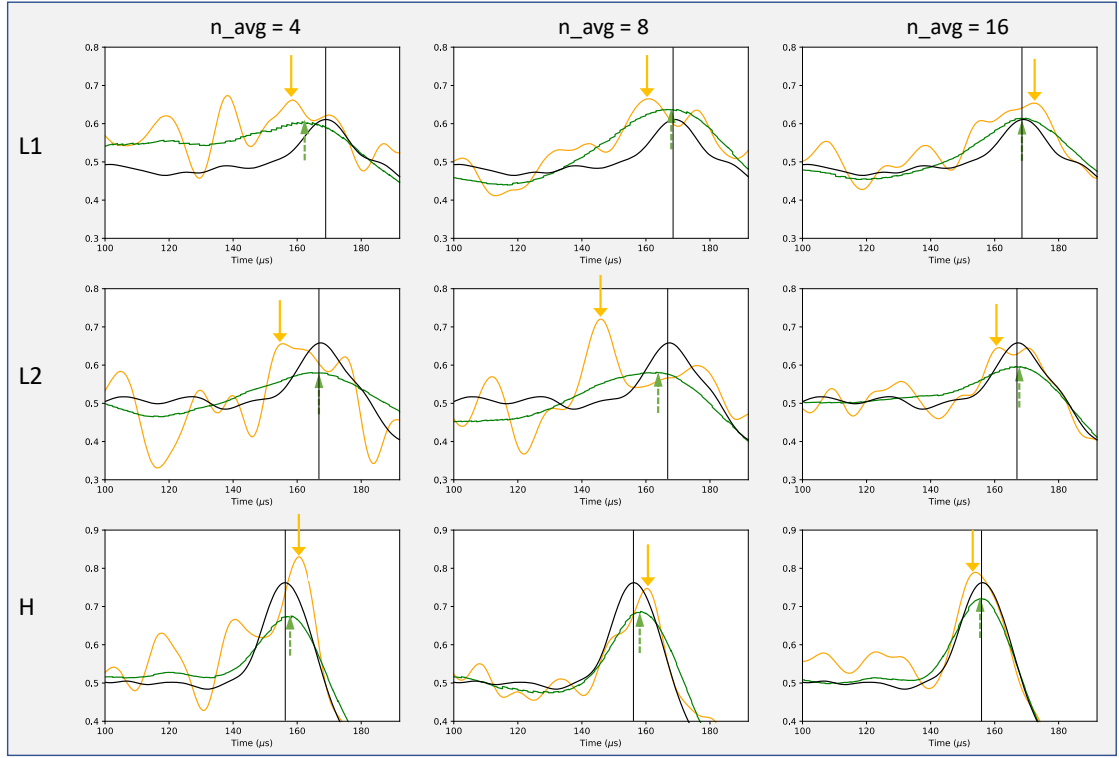

**Figure 5**. Examples of improved BP locations after SAE_clean denoising at n_avg = 4, 8, and 16, on all three detectors L1, L2, and H, zooming to the range between 100 and 192 μs. The orange arrows denote the BP position on noisy signals (orange curves) while green arrows denote for the denoised BP position. The black vertical lines denote the baseline BP position from the grand GT (black curves). Green arrows are closer to the black baseline compared to the orange arrows, showing smaller BP range uncertainties.

**Table 3**. The mean absolute error ($MAE^{BP}$) and mean error ($ME^{BP}$) representing BP range uncertainties before and after denoising, with n_avg >= 4.

| Experiment# | $MAE^{BP} \pm$ Std (mm) Before | $MAE^{BP} \pm$ Std (mm) After | $ME^{BP} \pm$ Std (mm) Before | $ME^{BP} \pm$ Std (mm) After |
|---|---|---|---|---|
| clean-L1-4 | 55.90 ± 70.87 | 41.00 ± 62.20 | -48.82 ± 75.92 | -32.52 ± 67.03 |
| clean-L1-8 | 32.08 ± 57.96 | 6.32 ± 8.41 | -25.40 ± 61.19 | -0.45 ± 8.41 |
| clean-L1-16 | 16.04 ± 36.38 | 5.41 ± 3.80 | -12.09 ± 37.88 | 1.44 ± 6.45 |
| clean-L1-24 | 5.13 ± 4.27 | 4.48 ± 3.21 | -0.32 ± 6.67 | 1.06 ± 5.41 |
| clean-L2-4 | 51.15 ± 70.02 | 28.95 ± 57.53 | -43.87 ± 74.80 | -18.48 ± 61.70 |
| clean-L2-8 | 27.45 ± 52.72 | 12.95 ± 35.35 | -22.27 ± 55.10 | -5.48 ± 37.25 |
| clean-L2-16 | 9.28 ± 24.77 | 3.95 ± 2.88 | -6.48 ± 25.65 | -0.23 ± 4.88 |
| clean-L2-24 | 5.75 ± 13.30 | 3.32 ± 2.55 | -0.66 ± 14.48 | 0.50 ± 4.16 |
| clean-H-4 | 16.22 ± 38.54 | 4.60 ± 3.16 | -10.93 ± 40.36 | 0.68 ± 5.54 |
| clean-H-8 | 6.74 ± 21.60 | 2.84 ± 1.95 | -2.12 ± 22.52 | 0.20 ± 3.44 |
| clean-H-16 | 3.15 ± 2.31 | 2.10 ± 1.78 | -0.31 ± 3.90 | 0.46 ± 2.71 |
| clean-H-24 | 2.30 ± 1.89 | 1.86 ± 1.59 | -0.11 ± 2.98 | -0.64 ± 2.36 |
| self-L1-4 | 39.97 ± 59.81 | 33.75 ± 54.82 | -30.64 ± 65.00 | -23.73 ± 59.84 |
| self-L1-8 | 33.77 ± 58.52 | 31.59 ± 60.45 | -28.20 ± 61.39 | -26.30 ± 62.93 |
| self-L1-16 | 12.24 ± 28.00 | 9.63 ± 23.96 | -5.84 ± 30.00 | -1.91 ± 25.75 |
| self-L1-24 | 6.43 ± 14.27 | 4.26 ± 3.32 | -3.12 ± 15.34 | -0.14 ± 5.40 |
| self -L2-4 | 43.05 ± 64.22 | 40.86 ± 67.31 | -33.41 ± 68.73 | -33.67 ± 71.18 |
| self -L2-8 | 30.33 ± 56.63 | 15.22 ± 41.37 | -23.29 ± 59.88 | -8.74 ± 43.20 |
| self -L2-16 | 6.38 ± 13.34 | 4.02 ± 3.14 | -0.86 ± 14.77 | 1.17 ± 4.98 |
| self -L2-24 | 3.67 ± 3.30 | 3.52 ± 2.55 | -0.88 ± 4.86 | 1.06 ± 4.22 |
| self -H-4 | 13.60 ± 35.77 | 6.50 ± 14.03 | -8.37 ± 37.34 | -1.91 ± 15.34 |
| self -H-8 | 6.43 ± 21.69 | 3.12 ± 2.40 | -1.71 ± 22.56 | 0.07 ± 3.93 |
| self -H-16 | 3.24 ± 2.35 | 1.99 ± 1.73 | -0.40 ± 3.99 | 0.32 ± 2.62 |
| self -H-24 | 2.43 ± 1.99 | 1.93 ± 1.14 | 0.10 ± 3.14 | 0.11 ± 2.24 |

## 4. Discussion

Protoacoustic method has been actively investigated for its feasibility of *in vivo* BP range verification. Though the method has unique advantages such as relatively simple instrumentation setup and straightforward correlation between the pressure/acoustic wave and BP localization, the key measuring objectives, acoustic signals, are susceptible to various sources of noises (thermal, scattering, and internal electric (Assmann *et al.*, 2015; Ahmad *et al.*, 2015; Jones *et al.*, 2016)), which in turn requires a large number of measures to achieve high SNRs. Such numerous measurements result in high dose delivery in tissue and not applicable for clinical use. We have proposed a SAE network based deep learning denoising method to reconstruct clean signals from only a small number of measurements, as well as to optimize the BP range verification by reducing the localization uncertainty.

We have tested two training strategies with SAE networks, namely the supervised SAE_clean and the self-supervised SAE_self. Overall, both models achieved substantial reduction of MSE and enhancement of SNR for low n_avg and essentially provided high quality denoised signals. From Table 2, for SAE_clean, the MSE reduced by about 80% at n_avg of 1 and about 50-60% at n_avg of 24, the SNR increased by 60-85% at n_avg of 1 and 10-20% at n_avg of 24, showing a steadily weaker enhancement of the signals for larger n_avg. This trending correlates to the fact that using more raw waves for average yields more stable and accurate signals. However, this trending is almost reversed for SAE_self models. According to Table 2, the MSE reduction and SNR increase were worse for low n_avgs of 1, 2, and 4, but better for high n_avgs of 8, 16, and 24. Such differences between SAE_clean and SAE_self models arise from the training philosophy behind them.

SAE as a successful denoising tool is often used for self-supervised (unsupervised learning), without any reference labels (Vincent *et al.*, 2010; Lin *et al.*, 2019; Huang and Sun, 2016; Thirukovalluru *et al.*, 2016).

SAE_self tries to map back to the input signals rather than any given ground truth. As shown in Figure 4, when n_avg equals to 1, the input signal was very noisy and lacking any obvious mode, SAE_self had no knowledge of underlying patterns and would simply treat those noises as some embedded truth to mimic during denoising, producing an output resembling those noises. But for SAE_clean, it is a supervised learning strategy that knows the clean ground truth and thus was able to distinguish those noisy components in low n_avg signals. As n_avg increased from 1 to 16, the input signals began to contain more information itself and show obvious underlying patterns, so SAE_self was also able to capture representative features apart from the random noises, and both SAE_self and SAE_clean achieved similar results. In many SAE denoising applications, the ground truth labels are missing or unavailable, but in this specific denoising task, we were lucky to have the averaged signals serving as ground truth and thus were able to change the training strategy from self-supervised learning to fully supervised learning, and achieved better results especially for small n_avgs.

One of the most important tasks in protoacoustic measurement is to localize BP in tissues. From Table 3, we see that the BP range uncertainty represented by $MAE^{BP}$ and $ME^{BP}$ was substantially reduced after SAE denoising, especially the standard deviations of $MAE^{BP}$ and $ME^{BP}$. Comparing the three detectors, overall, the high accuracy detector (H) performed better than the two low accuracy detectors (L1 and L2), both before and after denoising. For the high accuracy detector, the $ME^{BP}$ decreased from -10.93 ± 40.36 mm to 0.68 ± 5.54 mm by averaging only 4 raw signals, and decrease from -2.12 ± 22.52 to 0.20 ± 3.44 mm if averaging over 8 raw signals, while for L1 and L2, the uncertainty at n_avg = 8 are well above 5 mm. For L1, the $ME^{BP}$ decreased from -12.09 ± 37.88 mm to 1.44 ± 6.45 mm by averaging 16 raw signals. For L2, the $ME^{BP}$ decreased from -6.48 ± 25.65 to -0.23 ± 4.88 mm by averaging 16 raw signals. Such dramatic improvement suggests that with the aid of deep learning techniques, it is possible to achieve accurate BP range verification with just a few acoustic signals, which brings down the dose and time required in protoacoustic measurements.

Previous literatures have reported BP range uncertainty in both simulation and experimental studies. The reported BP range uncertainty varied from submillimeter to 5 millimeters (Otero *et al.*, 2019; Paganetti, 2012; Kellnberger *et al.*, 2016; Assmann *et al.*, 2015; Jones *et al.*, 2018), and in some extreme cases, the standard deviation was up to 10 mm (Paganetti, 2012). Moreover, Jones *et al.* found that the BP range uncertainty depends on the proton pulse width (Jones *et al.*, 2016). For an extremely narrow Dirac-delta-function-like (FWHM < 4 μs) proton pulse, the systematic error of BP determination is <2.6 mm, but for longer non-δ-function-like beams (FWHM = 56 μs), a systematic error up to 23 mm can be expected. Narrower proton pulse favors in producing acoustic pressure waves; however, the amplitude of acoustic signal is also limited by the energy deposited by proton pulse, assuming the peak proton beam current remains at the same level. Therefore, the desired proton pulse width for typical medical proton cyclotron is ~10-14μs, the proton beam used in our was modulated by a function generator, where ~14-18μs pulse width (FWHM) was achieved. the expected BP range uncertainty should be around a few millimeters, much greater than the optimal submillimeter records (Assmann *et al.*, 2015), which is consistent with our denoised results. When treat patient with proton beams, typically a 3.5% of the range plus 1-3 mm margin is allowed to account for the uncertainty of BP range (Paganetti, 2012), corresponding to an uncertainty of 3-5 mm for the BP depth of around 50-60 mm in our experiment. Our denoised results of the high accuracy detector fall well within that range and the low accuracy detectors fall marginally in that range.

The main idea of this work is to reduce the dose delivery without compromising the accuracy of BP range verification. In our measurement, each raw acoustic signal was generated by a proton pulse equivalent to 2.36 cGy, so the dose delivery would be 24.2 Gy to average 512 signals. After denoising, we can obtain BP localization with 16-24 signals for the two low sensitive detectors and with only 4-8 signals for the high accuracy detector, corresponding to 37.8-56.6 cGy and 9.4-18.9 cGy, respectively. The required doses for BP range verification were largely reduced with the aid of deep learning denoising techniques.

Our study is performed with the proton beam of one specified energy of 226 MeV, incident on a uniform PE phantom, while in clinics the patient tissues are heterogeneous and behave more complicated than the phantom. Though we tested our SAE denoising method on a relatively simple setup, to the best of our knowledge, this is the first trial of applying deep learning SAE models to denoise the protoacoustic signals, and achieved satisfactory results. However, as we have discussed in the above paragraphs, the proton beam we used was not δ-function-like pulses and could intrinsically increase the BP range uncertainty. Besides, the high accuracy detector outperformed the low accuracy ones in all metrics. Other techniques, such as acoustic array with time-reversal algorithm, can be used to further reduce the BP range uncertainties, but it is beyond the scope of this study (Yu *et al.*, 2021). To push for better results in future, possible solutions include using narrower proton pulses as the dose deposition source and sticking to the higher accuracy acoustic wave detectors available.

## 5. Conclusion

We have demonstrated that the SAE networks can be used to denoise protoacoustic signals for BP range verification. Besides the commonly used self-supervised training strategy, we introduced the full supervised learning using the averaged signals (averaging 192 raw signals) as ground truth and achieved better results. Decreased MSEs and increased SNRs were obtained for all experiments after SAE denoising. For the high accuracy detector, denoised BP uncertainty was well within 5 mm by averaging only 8 raw signals. For the low accuracy detectors, denoised BP uncertainty was marginally within 5 mm by averaging 16-24 raw signals. Deep learning denoising techniques can be integrated to the data acquisition for signal processing during protoacoustic measurements, to reduce the dose required for obtaining stable and clean signals.

## Conflict(s) of interest

None.

## Acknowledgements

This work is supported in part by the National Institutes of Health under Award Number R01CA215718 and R56EB033332.